\documentclass{article}

\usepackage{PRIMEarxiv}

\usepackage[utf8]{inputenc} 
\usepackage[T1]{fontenc}    
\usepackage{hyperref}       
\usepackage{url}            
\usepackage{booktabs}       
\usepackage{amsfonts}       
\usepackage{nicefrac}       
\usepackage{microtype}      
\usepackage{lipsum}
\usepackage{fancyhdr}       
\usepackage{graphicx}       
\graphicspath{{media/}}     
\usepackage{authblk}

\title{Digital Health Discussion Through Articles Published Until the Year 2021: A Digital Topic Modeling Approach
}

\author[1]{\small Junhyoun Sung}
\author[2, 3, 4]{\small Hyungsook Kim}

\affil[1]{\footnotesize Department of Biostatistics, University of Washington, Seattle, WA, USA}
\affil[2]{\footnotesize Department of Cognitive Sciences, School of Intelligence, Hanyang University, Seoul, Republic of Korea}
\affil[3]{\footnotesize Graduate School of Public Policy, Hanyang University, Seoul, Republic of Korea}
\affil[4]{\footnotesize HY Digital Healthcare Center, Hanyang University, Seoul, Republic of Korea}

\begin{document}
\maketitle

\begin{abstract}
\item The digital health industry has grown in popularity since the 2010s, but there has been limited analysis of the topics discussed in the field across academic disciplines. This study aims to analyze the research trends of digital health-related articles published on the Web of Science until 2021, in order to understand the concentration, scope, and characteristics of the research. 15,950 digital health-related papers from the top 10 academic fields were analyzed using the Web of Science. The papers were grouped into three domains: public health, medicine, and electrical engineering and computer science (EECS). Two time periods (2012-2016 and 2017-2021) were compared using Latent Dirichlet Allocation (LDA) for topic modeling. The number of topics was determined based on coherence score, and topic compositions were compared using a homogeneity test. The number of optimal topics varied across domains and time periods. For public health, the first and second halves had 13 and 19 topics, respectively. Medicine had 14 and 25 topics, and EECS had 7 and 21 topics. Text analysis revealed shared topics among the domains, but with variations in composition. The homogeneity test confirmed significant differences between the groups (adjusted p-value<0.05). Six dominant themes emerged, including journal article methodology, information technology, medical issues, population demographics, social phenomena, and healthcare. Digital health research is expanding and evolving, particularly in relation to Covid-19, where topics such as depression and mental disorders, education, and physical activity have gained prominence. There was no bias in topic composition among the three domains, but other fields like kinesiology or psychology could contribute to future digital health research. Exploring expanded topics that reflect people's needs for digital health over time will be crucial.
\end{abstract}


\section{Introduction}

Public interest in digital health in the international community has soared since the 2010s \cite{ref1}. Terms such as “digital platform” and “digital therapy” began to emerge in the 2000s, but digital health is still considered a special area \cite{ref2}.  However, with the expansion of digital techniques, healthcare programs of platform companies have recently gained ground \cite{ref3}. In addition to the recorded content, some programs provide feedback or health-related counseling in real time \cite{ref4}. For instance, wearable sensing gloves and sensory feedback advance the quality of life for people with impaired limbs \cite{ref5}.  

Digital health is not just a concept limited to cure or exercise but is itself a value. It is an industry that includes the so-called healthy life, and wellness, as well as perspectives on therapeutics and daily care \cite{ref6}. The definition of health has a vast range, including exercise and diet, body and mind, and prevention and management of treatment and cure. With the development of the digital market, health has grown into a concept that includes various interactions and content in a non-face-to-face environment after the outbreak of Covid-19 \cite{ref7}.

On the other hand, as the concept expands, it is necessary to analyze whether research topics related to digital health have also been expanded. Research trend analysis belongs to the field of Scientometrics. It is a methodology that quantitatively analyzes the contents of documents, such as abstracts, citation relationships between documents, and the characteristics of authors through statistical techniques \cite{ref8}. The advantage of Scientometrics is that it can objectively and quantitatively measure the importance of a research topic through the frequency of keywords, and it is easy to grasp the size and history of the academic field to which the thesis belongs \cite{ref9}. In the global information age, it is vital to quickly and accurately secure the information we need, so it is meaningful in terms of efficiency to analyze the abstract that notifies and criticizes the content of the thesis as an information retrieval tool \cite{ref10}. In particular, if topic modeling is applied to the abstract, it is possible to comprehend all the words and messages appearing in the corresponding message by organizing the contents by topic \cite{ref11}. 

As such, the concept of digital health continues to expand with the growing associated industries. However, previous studies that have examined the changes and trends in digital health are scarce. This study compares and analyzes the distribution of published articles focusing on digital health. The number of articles is investigated by academic field, domain, and time period. More specifically, this paper looks into what topics are being studied and what are the dominant themes. It compares each domain’s perspective on digital health and checks, as well as changes in the composition of highlighted topics, if any, in each academic field over time by applying statistical testing. 

It is presumed that most articles published focusing on digital health belong to the public health or computer science field and change in theme composition, and this will be verified in this paper. In addition, this article intends to examine whether the overall digital health-related research topic has been expanded by applying text analysis and whether most articles focus on a specific topic, and then suggest the necessity of expanded digital health research.

\section{Methods}
\label{sec:headings}

\subsection{Terms and Definitions}
Terms need to be defined to understand this paper. Since this article focuses on trends, many terms such as 'fields,' 'topics,' 'themes,' and 'domains’ will be used. The definition of each term is shown in Table 1. This way of classification is applied in many topic modeling articles \cite{ref12, ref13, ref14}.

\begin{table}[hbt!]
 \caption{Definition of terms in use}
  \centering
  \begin{tabular}{lll}
    \cmidrule(r){1-2}
    Name     & Description \\
    \midrule
    Academic Field & The field of study. Majors, concentrations, or specializations.               \\
    Domain         & The umbrella term of academic fields. The domain consists of academic fields. \\
    Topic & The topic of abstracts. The topic of a group of papers.    \\
    Theme & The umbrella term of topics. The theme consists of topics. \\
    \bottomrule
  \end{tabular}
  \label{tab:table1}
\end{table}

\subsection{Data Collection}
For the collection of data in this study, articles related to digital health published on the Web of Science(WoS) until 2021 were searched and crawled on January 15, 2022. WoS is a database for the Science Citation Index Expanded(SCIE) articles which are high quality, so it represents the trends of digital health more strictly. Although there are many databases, this paper will investigate only SCI papers this time, and other papers could be analyzed in later research. 

In the detailed search, only papers with the theme (headings, abstracts, and keywords) “Digital Health” were searched and 27,638 articles were extracted. These articles belonged to 239 academic fields; among them, the top ten academic fields were on healthcare sciences and services, public, environmental, and occupational health, medical informatics, electrical and electronics engineering , computer science and information systems, general internal medicine, computer science, interdisciplinary applications, psychiatry, computer science theory  and methods, and health policy and services. These topics were grouped into three domains: public health, medicine, and electrical engineering and computer science (EECS). While 18,704 papers were in the top ten categories, only 15,950 papers remained after excluding papers with overlapping categories. This study analyzed the English abstracts of those papers.

\subsection{Modeling}
This study applied text mining, particularly topic modeling, to analyze the overall research trends in the digital health field. Text mining refers to the overall process of analyzing the frequency of word appearance, degree of connection between words, and relationship between multiple word units for data composed of strings \cite{ref15}. 

Frequency analysis, which is the basis of text mining, can be performed by focusing on word count or publication year. Overall, it is a good analysis method for verifying which words are frequently mentioned and how the distribution of words varies over time \cite{ref16}.

Topic analysis refers to the clustering and classification of multiple words in the entire document beyond understanding the text in word units. Topic modeling was first proposed as an algorithm to determine topics that exist in a set of documents \cite{ref17}. Topic analysis is unsupervised learning. Based on the correlation between words, groups are determined in the direction of increasing heterogeneity between clusters and increasing homogeneity within groups. Because the number of topics determines the results of topic modeling, a reasonable number of topics must be determined in the entire document set \cite{ref18}. The optimal number of topics should be determined through trial and error to avoid overlapping topics and ensure the individuality of topics \cite{ref19}. Contrarily, some argue that the smaller the number of topics, the easier it is to ensure structural effectiveness, because duplication can be avoided \cite{ref20}. In this study, the optimal number of topics was determined using a coherence score \cite{ref21}.

There are various methodologies for topic analysis, and latent Dirichlet allocation (LDA) exists as a representative method. The LDA proposed by Blei et al. assumes that topics and words in the topic in the document follow the Dirichlet distribution \cite{ref22}. LDA is a probability distribution model that extracts potentially significant topics from multiple documents based on the simultaneous appearance of documents and allocates words corresponding to each topic by calculating parameters such as the quantity, number of words, and corpus \cite{ref23}. LDA selects a topic from the Dirichlet distribution after observing the words included in the document. It then adds words belonging to the topic and moves on to the next topic when there is no such word. This process is repeated to cluster words \cite{ref24}. However, because a word across multiple topics cannot be selected for another topic when it is selected for any topic, bias exists according to the initial selection \cite{ref25}. Since the initial selection is random, the clustering contents may vary when LDA is repeated \cite{ref26}. Because LDA is unsupervised learning, it is not free from the problem of labeling clusters \cite{ref27}.

Lambda values can be adjusted in the process of visualizing LDA, which is the probability of terms appearing on a specific topic divided by the probability of words appearing throughout a document. In short, the closer lambda is to 1, the more likely it is that words appear throughout the document; the closer lambda is to 0, the more the likelihood of prominent words being expressed in a specific document \cite{ref28}. The analyst should find the optimal interpretation by changing lambda to a value between 0 and 1. If lambda is close to 0, the characteristics of the subject are emphasized, but unnecessary junk words may also be extracted; if lambda is close to 1, words that reveal the characteristics of the subject may not appear \cite{ref29}.

After clustering the subjects through LDA, this study examines whether the cluster composition changes over time. The articles from 2012 to 2016 were assigned to the first half, and the articles from 2017 to 2021 are assigned to the second half to determine the changes in topic composition by time period. The Fisher's exact test is employed to assess the independence of the topic distributions across different categories in the first and second halves, providing a robust solution especially for small sample sizes \cite{ref30}. 

\subsection{Processing}
For the 15950 papers extracted from the Web of Science, text mining was conducted in general, by applying topic analysis using Python. First, in the preprocessing stage, terminology such as “thing” and “etc.”  as well as meaningless words of three or less alphabets were removed. Thereafter morpheme analysis was conducted.

The reason for choosing LDA as a topic-modeling technique is that the LDA assumes that one document may have several types of topics. LDA can grasp the subject of a paper more complexly than K-means clustering, which only searches for one topic \cite{ref31}. Prior to proceeding with LDA, the coherence score was used in Python to optimize the number of topics.

After conducting LDA analysis at each academic field level, this study implemented the independence test using R to check the differences in topic composition by period and academic field. This process is visualized as a diagram in Figure 1. 

\begin{figure}[hbt!]
  \centering
  \includegraphics[width=0.9\textwidth]{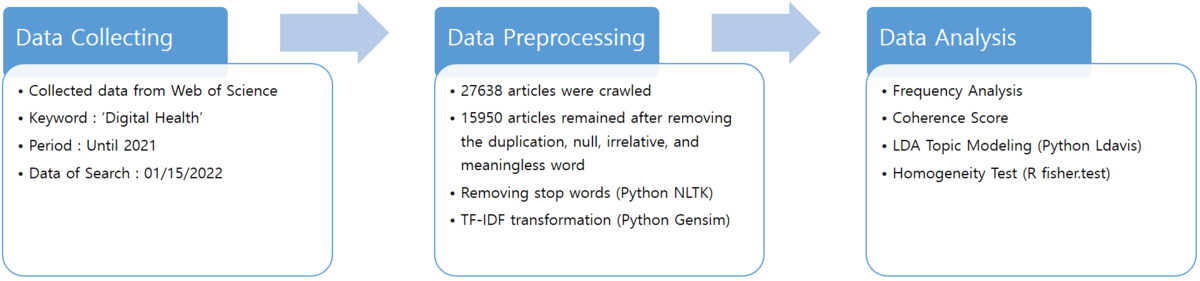}
  \caption{Sample Flow chart showing analysis procedure}
  \label{fig:figure1}
\end{figure}

\section{Results}
\label{sec:results}
\subsection{Time-series Analysis}
The Table 2 shows the number of papers published annually on digital health and indicates an overall rise from 2012 to 2021. Approximately, 1,000 more papers were published at the time of the Covid-19 outbreak in 2019 as compared to 2018, and more than 5,000 papers were published annually in 2020 and 2021. The number of papers published in 2012 was 557; it increased by more than 10 times by 2021.

This study aims to compare the composition of topics by period and understand the changes. The period between 2012 and 2016 was considered the first half for the publishing of articles, while the period between 2017 and 2021 was considered the second half.

The scope of digital health has continued to expand as the number of related academic papers has increased over the last decade. Table 1 also shows that almost 200 academic fields have been related to digital health since 2012.

\begin{table}[hbt!]
 \caption{The number of articles published on digital health and academic fields tagged as digital health}
\centering
\begin{tabular}{ccccccccccc}
\hline
Year                              & 2012 & 2013 & 2014 & 2015 & 2016 & 2017 & 2018 & 2019 & 2020 & 2021 \\ \hline
Number of digital health articles & 557  & 734  & 789  & 1064 & 1352 & 1916 & 2580 & 3459 & 5021 & 6174 \\
Number of academic fields         & 199  & 205  & 209  & 219  & 217  & 226  & 230  & 232  & 235  & 239  \\ \hline
\end{tabular}
\label{tab:table2}
\end{table}

Of the 27,638 papers published until 2021, the total number of papers from the top ten fields were 18,704 (Figure 2). The top ten fields can be divided into three domains and the results show the numbers after the removal of duplication and missing publication years (Figure 3). This study only focuses on 15,950 papers belonging to the top ten fields. The distribution of articles excluding iterations and number of articles is shown in Figure 3, using a Venn diagram. Since all articles have academic field tags, articles belong to several fields are assigned to overlapped areas in a Venn diagram.

We asked medical doctors and professors in Computer Science and Public Health department to verify the validity of this reclassification. The reclassification was done based on considering the contents of the fields of study. This would help people comprehend the entire trend from a macroscopic point of view rather than analyzing every field of study.

\begin{figure}[hbt!]
  \centering
  \includegraphics[width=0.65\textwidth]{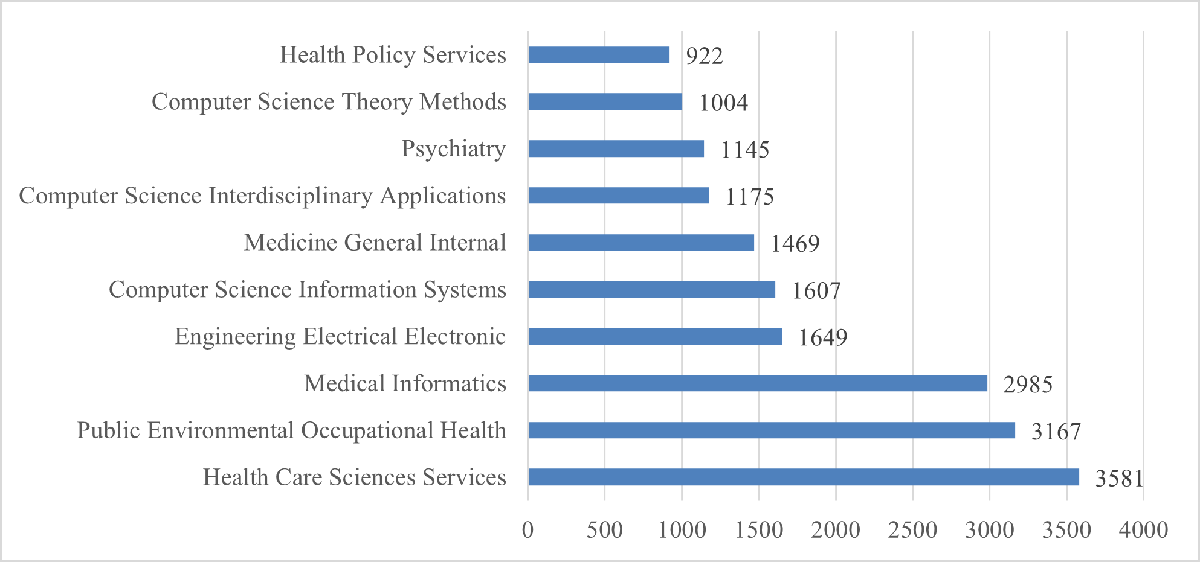}
  \caption{Top ten fields related to digital health from the year 1955 to the year 2021}
  \label{fig:figure2}
\end{figure}

\begin{figure}[hbt!]
  \centering
  \includegraphics[width=0.7\textwidth]{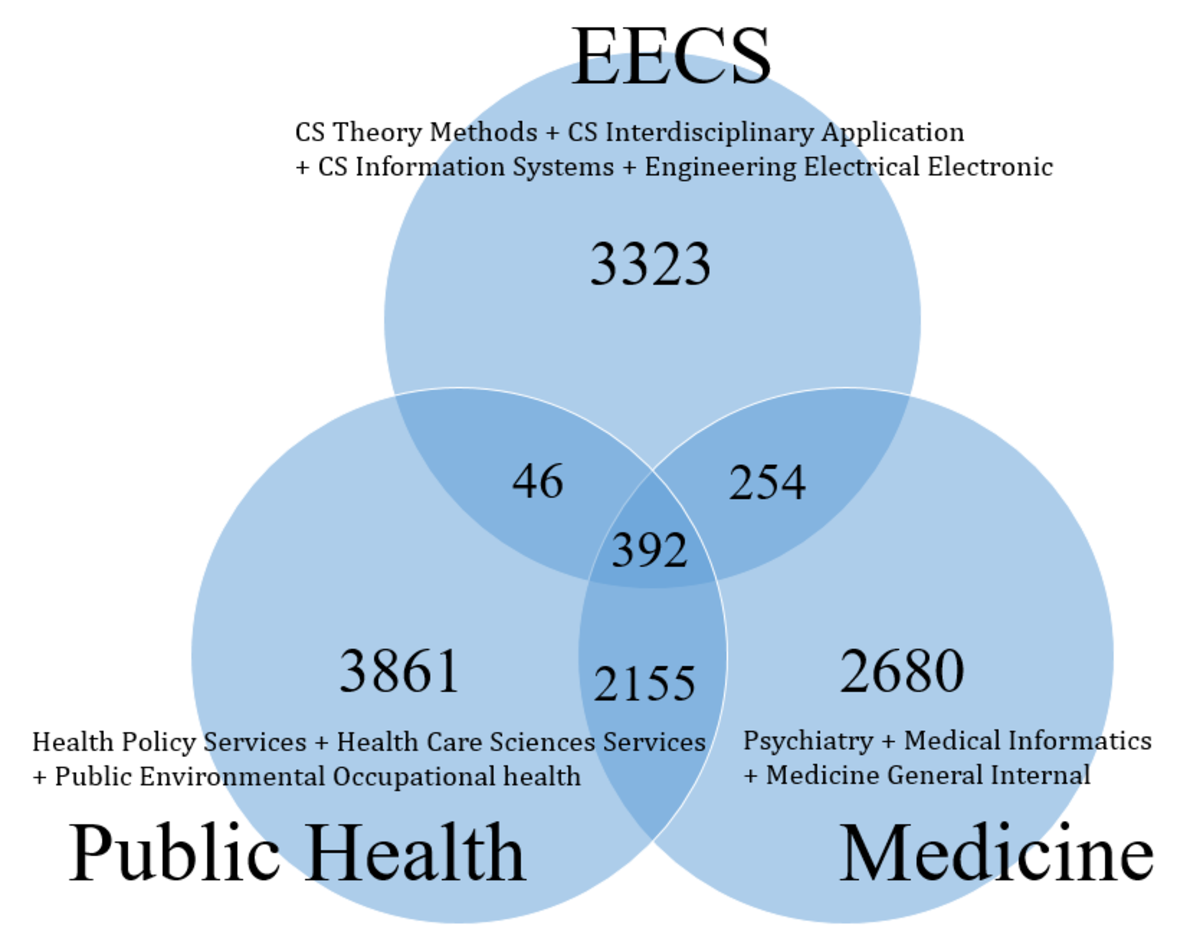}
  \caption{Number of articles distributed among three domains until 2021}
  \label{fig:figure3}
\end{figure}

\subsection{Topic Number Selection}
The coherence scores were calculated using Python to select the optimal number of topics. They were calculated by increasing the number of topics in one step from 2 to 40; the optimal number of topics determined for each period and academic field is shown in Figure 4 and Table 3.

\begin{figure}[hbt!]
  \centering
  \includegraphics[width=\textwidth]{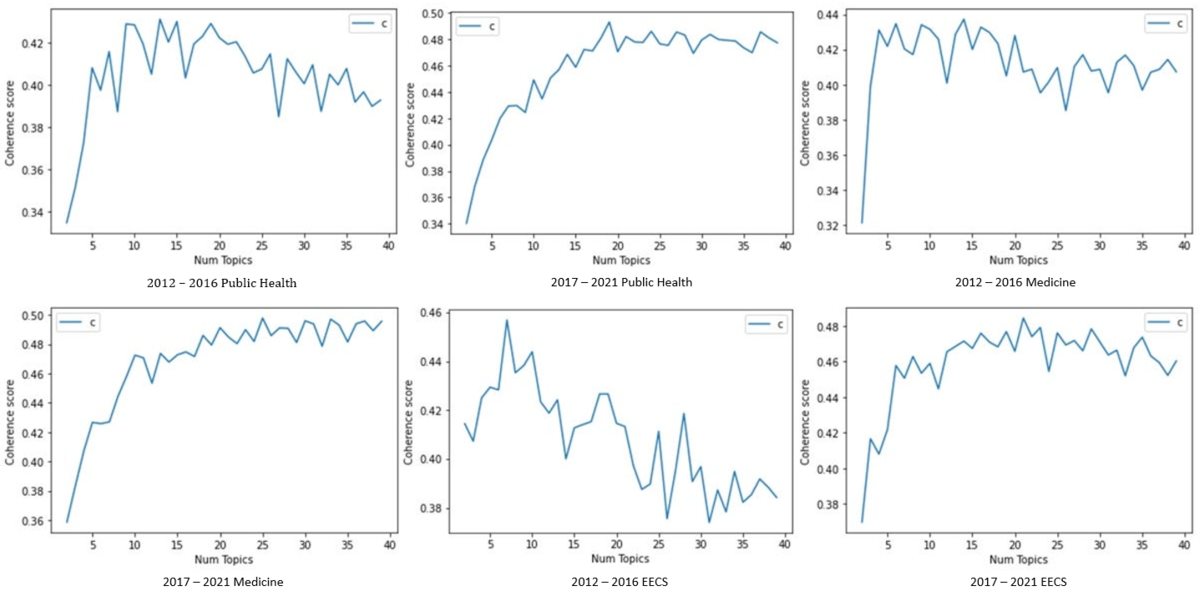}
  \caption{Best coherence scores for each topic}
  \label{fig:figure4}
\end{figure}

\begin{table}[hbt!]
 \caption{Definition of terms in use}
  \centering
  \begin{tabular}{lll}
    \cmidrule(r){1-3}
    Category     & Number of Topics & Coherence Score \\
    \midrule
    2012 - 2016 Public Health & 13 & 0.43 \\
    2017 - 2021 Public Health & 19 & 0.49 \\
    2012 - 2016 Medicine & 14 & 0.44 \\
    2017 - 2021 Medicine & 25 & 0.50 \\
    2012 - 2016 EECS & 7 & 0.46 \\
    2017 - 2021 EECS & 21 & 0.48 \\
    \bottomrule
  \end{tabular}
  \label{tab:table3}
\end{table}

\subsection{Latent Dirichlet Allocation}
There was a total of six LDA topic modeling results according to period and academic field. The names of the LDA topic models are listed in Table 2. Each model had its own topic distribution. For instance, Figure 5 shows the LDA topic model of the topics of public health from 2012 to 2016. The number of topics was determined to be 13, and the 30 words that mostly stood out in topic 1 appear on the right side of Figure 5. The set of the 30 most prominent words varied as the lambda value changed. In this study, lambda was applied consistently at 0.6 in every LDA topic model because the lambda value has the highest explanatory power when it is 0.6 \cite{ref32}.

The Principal Component1 (PC1) showed practicality. The more the topic is on the right-hand side, the more theoretical it is. PC2 showed macroscopic characteristics.  The more the topic is at the top, the more specific the topic is.

\begin{figure}[hbt!]
  \centering
  \includegraphics[width=\textwidth]{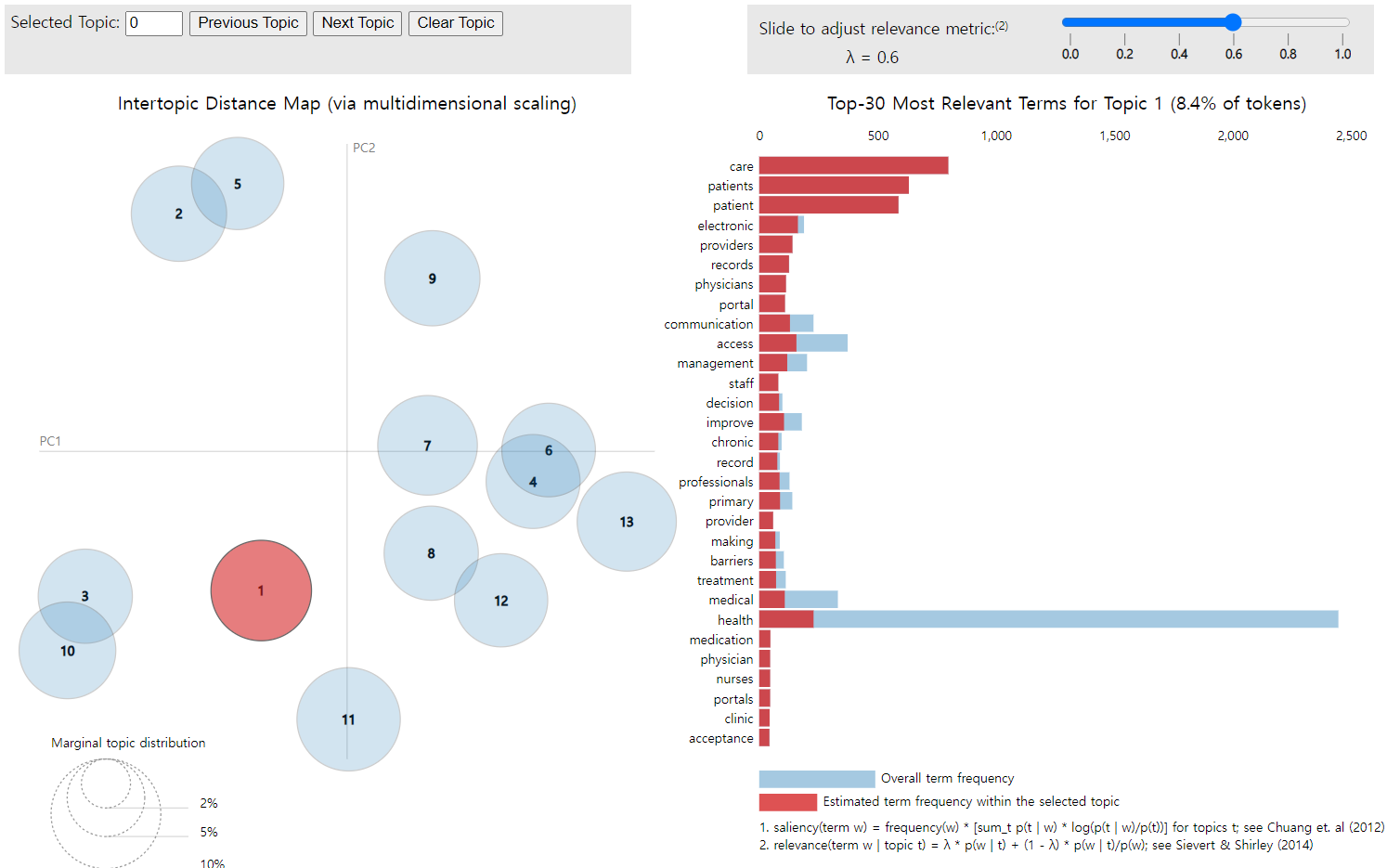}
  \caption{LDA clustering for public health from 2012 to 2016}
  \label{fig:figure5}
\end{figure}

After classifying topics that appeared in the public health field from 2012 to 2016, these topics were clustered into each theme. The themes in bold represent a collection of topics that are similar in content. The top 30 most relevant terms for each topic were closely examined. Topics located in close proximity are not always clustered as a group, because PC1 and PC2 represent only partial characteristics. Sometimes, different topics share some characteristics, while similar topics do not; hence, theme clustering was executed based on text analysis. Table 4 lists the number of articles and their topic distributions.

\begin{table}[hbt!]
 \caption{Topic distribution of public health from 2012 to 2016}
\centering
\begin{tabular}{ll}
\hline 

Theme,   Topics, and Keywords                                                                                     & Articles   (N=813), n (\%) \\ \hline
\\ \textbf{Theme 1: Journal Article Methodology}                                                                     &                            \\  \\
\begin{tabular}[c]{@{}l@{}}- Topic   13: Group Experiment\\ - Keywords:   Group, program, control, trial, randomized, week, follow, recruited\end{tabular} &
  59 (7.3\%) \\
                                                                                                                  &                            \\
\textbf{Theme 2: Information Technology}                                                                          &                            \\  \\
\begin{tabular}[c]{@{}l@{}}- Topic   12: Mobile\\     - Keywords:   Mobile, apps, devices, ehealth, mhealth, device, phones, smartphones\end{tabular} &
  63 (7.8\%) \\ \\
\begin{tabular}[c]{@{}l@{}}- Topic   8: Game\\   - Keywords:   Game, content, video, play, engage, interactive\end{tabular} &
  54 (6.6\%) \\  \\
\begin{tabular}[c]{@{}l@{}}- Topic   7: System\\   - Keywords:   systematic, databases\end{tabular}           & 61 (7.5\%)                 \\   \\
\begin{tabular}[c]{@{}l@{}}- Topic   2: Framework \& Architecture\\   - Keywords:   environment, framework, resources\end{tabular} &
  70 (8.6\%) \\  \\
\begin{tabular}[c]{@{}l@{}}- Topic   10: Privacy \& Security\\     - Keywords:   technology, healthcare, privacy, security\end{tabular} &
  80 (9.8\%) \\  \\
                                                                                                                                         
\textbf{Theme 3: Medical Issues}                                                                                  &                            \\  \\
\begin{tabular}[c]{@{}l@{}}- Topic   6: Cancer\\  - Keywords:   Cancer, risk, breast, mortality\end{tabular} & 52 (6.4\%)                 \\   \\
\begin{tabular}[c]{@{}l@{}}- Topic   5: Hospital\\    - Keywords:   Clinical, medical, telemedicine, dental, hospital, medicine\end{tabular} &
  60 (7.4\%) \\  \\
\begin{tabular}[c]{@{}l@{}}- Topic   4: Blood\\   - Keywords:   blood, dose, heart\end{tabular}               & 57 (7.0\%)                 \\   \\
\begin{tabular}[c]{@{}l@{}}- Topic   1: Clinical Consultation\\     - Keywords:   medical, medication, nurses, clinic\end{tabular} &
  76 (9.3\%) \\ 
                                                                                                                  &                            \\
\textbf{Theme 4: Population Demographics}                                                                         &                            \\  \\
\begin{tabular}[c]{@{}l@{}}- Topic   3: Youth\\    - Keywords:   Youth, students\end{tabular}                  & 60 (7.4\%)                 \\  \\
\begin{tabular}[c]{@{}l@{}}- Topic   11: Older\\   - Keywords:   adults, older, users, family\end{tabular}    & 64 (7.9\%)                 \\ 
                                                                                                                  &                            \\
\textbf{Theme 5: Social Phenomena}                                                                                &                            \\   \\
\begin{tabular}[c]{@{}l@{}}- Topic   9: Area\\   - Keywords:   urban, state, area, public, national, local\end{tabular} &
  57 (7.0\%) \\   \hline
\end{tabular}
\label{tab:table4}
\end{table}

This type of classification can be applied to all the six LDA topic models. The integrated results are listed in Table 5. All topics remain the same in six themes: journal article methodology, IT, medical issues, population demographics, social phenomenon, and healthcare.

Table 6 lists the number of articles on each theme and topic. It shows the percentages of topic compositions so that the weight of each theme and topic can be established.

\begin{table}[!hbt]
\caption{Classification of topics and themes}
\centering
\begin{tabular}{lcccccc}
\cline{1-7}
 &
  \begin{tabular}[c]{@{}c@{}}Public health\\  12-16\end{tabular} &
  \begin{tabular}[c]{@{}c@{}}Public health\\   17-21\end{tabular} &
  \begin{tabular}[c]{@{}c@{}}Medicine\\   12-16\end{tabular} &
  \begin{tabular}[c]{@{}c@{}}Medicine\\   17-21\end{tabular} &
  \begin{tabular}[c]{@{}c@{}}EECS\\   12-16\end{tabular} &
  \begin{tabular}[c]{@{}c@{}}EECS\\  17-21\end{tabular} \\ \cline{1-7}
\textbf{1. Journal Article Methodology} &          &          &          &          &         &          \\
- Model Validation                      &          & Topic 5  & Topic 13 & Topic 5  &         & Topic 17 \\
- Group Experiment                      & Topic 13 & Topic 15 &          & Topic 8  &         &          \\
- Literature Review                     &          & Topic 11 & Topic 7  & Topic 22 &         & Topic 11 \\
- Quality Analysis                      &          &          &          & Topic 17 &         & Topic 21 \\
\textbf{2. Information Technology}      &          &          &          &          &         &          \\
- Mobile                                & Topic 12 & Topic 4  & Topic 3  & Topic 7  & Topic 6 & Topic 3  \\
- UI/UX                                 &          &          &          & Topic 1  &         &          \\
- Phenotyping                           &          &          &          & Topic 6  &         &          \\
- Game (Gamification, VR)               & Topic 8  &          & Topic 2  &          & Topic 4 & Topic 18 \\
- Platform                              &          &          &          &          & Topic 7 &          \\
- System                                & Topic 7  &          &          & Topic 25 &         & Topic 6  \\
- Framework \&   Architecture           & Topic 2  & Topic 9  &          &          &         & Topic 2  \\
- Privacy \& Security                   & Topic 10 & Topic 3  & Topic 10 &          & Topic 3 & Topic 4  \\
- Sensor Device                         &          &          & Topic 14 &          & Topic 2 & Topic 7  \\
- Image Identification                  &          & Topic 18 & Topic 1  &          & Topic 1 & Topic 12 \\
- Hardware \& Battery                   &          &          &          &          &         & Topic 8  \\
\textbf{3. Medical Issues}              &          &          &          &          &         &          \\
- Covid                                 &          & Topic 16 &          & Topic 13 &         & Topic 5  \\
- Cancer                                & Topic 6  &          & Topic 4  & Topic 11 &         &          \\
- Hospital (Cure)                       & Topic 5  &          &          &          &         & Topic 13 \\
- Blood                                 & Topic 4  &          &          &          &         & Topic 15 \\
- Smoking                               &          &          &          & Topic 3  &         &          \\
- Clinical Consultation                 & Topic 1  & Topic 14 & Topic 12 & Topic 15 &         &          \\
- Detecting Model                       &          &          &          & Topic 19 &         & Topic 19 \\
- Depression \& Mental   Disorder       &          & Topic 8  &          & Topic 10 &         & Topic 9  \\
\textbf{4. Population Demographics}     &          &          &          &          &         &          \\
- Family                                &          & Topic 1  &          &          &         &          \\
- Child                                 &          &          &          & Topic 14 &         &          \\
- Youth                                 & Topic 3  &          &          & Topic 16 &         &          \\
- Older                                 & Topic 11 & Topic 6  & Topic 8  &          &         &          \\
- Gender                                &          & Topic 2  &          & Topic 20 &         &          \\
\textbf{5. Social Phenomena}            &          &          &          &          &         &          \\
- Industry                              &          & Topic 13 &          &          &         & Topic 14 \\
- Area                                  & Topic 9  &          &          & Topic 9  &         &          \\
- Social Media                          &          &          & Topic 5  & Topic 24 &         & Topic 16 \\
- Community                             &          & Topic 17 &          & Topic 18 & Topic 5 &          \\
- Futurology                            &          &          & Topic 11 & Topic 4  &         & Topic 1  \\
- Online Communication                  &          & Topic 19 &          &          &         &          \\
- Education                             &          & Topic 7  &          & Topic 2  &         & Topic 20 \\
- Innovation \& Ethics                  &          &          &          & Topic 21 &         &          \\
\textbf{6. Healthcare}                  &          &          &          &          &         &          \\
- Physical Activity                     &          & Topic 12 & Topic 6  & Topic 12 &         & Topic 10 \\
- Daily Care (prevention)               &          & Topic 10 & Topic 9  & Topic 23 &         &          \\
\textbf{Total Number of Themes}         & \textbf{13}       & \textbf{19}       & \textbf{14}       & \textbf{25}       & \textbf{7}       & \textbf{21}      \\ \hline
\end{tabular}
\label{tab:table5}
\end{table}

Public health, in the first half, focused on IT (40\%) and medical issues (30\%). In the second half, papers on public health take a close look at social phenomena and begin to focus on healthcare. The total percentage of papers related to social phenomena and healthcare was 7, while it surged to 31 in the second half. The most dominant topics in the second half were community, online communication, education, physical activity, and daily care. The public health field is not clearly distinguished from medicine in the first half.  However, the topics in the second half highlighted social phenomena and daily care issues.

\begin{table}[!hbt]
\caption{The number and percentage of published articles according to topics and themes }
\centering
\begin{tabular}{lcccccc}
\hline
 &
  \begin{tabular}[c]{@{}c@{}}Public health\\   12-16\end{tabular} &
  \begin{tabular}[c]{@{}c@{}}Public health\\   17-21\end{tabular} &
  \begin{tabular}[c]{@{}c@{}}Medicine\\   12-16\end{tabular} &
  \begin{tabular}[c]{@{}c@{}}Medicine\\   17-21\end{tabular} &
  \begin{tabular}[c]{@{}c@{}}EECS\\    12-16\end{tabular} &
  \begin{tabular}[c]{@{}c@{}}EECS\\   17-21\end{tabular} \\ \hline
\textbf{1. Journal Article Methodology} & \textbf{59(7.3)}   & \textbf{850(18.8)}  & \textbf{73(11.8)}  & \textbf{824(20.9)} & \textbf{0(0.0)}    & \textbf{222(10.5)} \\
- Model Validation                      &                    & 198(4.4)            & 29(4.7)            & 116(2.9)           &                    & 75(3.5)            \\
- Group Experiment                      & 59(7.3)            & 273(6.0)            &                    & 153(3.9)           &                    &                    \\
- Literature Review                     &                    & 379(8.4)            & 44(7.1)            & 371(9.4)           &                    & 107(5.1)           \\
- Quality Analysis                      &                    &                     &                    & 184(4.7)           &                    & 40(1.9)            \\
\textbf{2. Information Technology}      & \textbf{328(40.3)} & \textbf{819(18.1)}  & \textbf{245(39.5)} & \textbf{629(16.0)} & \textbf{765(83.6)} & \textbf{964(45.6)} \\
- Mobile                                & 63(7.8)            & 134(3.0)            & 36(5.8)            & 142(3.6)           & 114(12.5)          & 54(2.6)            \\
- UI/UX                                 &                    &                     &                    & 152(3.9)           &                    &                    \\
- Phenotyping                           &                    &                     &                    & 164(4.2)           &                    &                    \\
- Games (Gamification, VR)              & 54(6.6)            &                     & 45(7.3)            &                    & 100(10.9)          & 123(5.8)           \\
- Platform                              &                    &                     &                    &                    & 88(9.6)            &                    \\
- System                                & 61(7.5)            &                     &                    & 171(4.3)           &                    & 111(5.2)           \\
- Framework \&   Architecture           & 70(8.6)            & 252(5.6)            &                    &                    &                    & 98(4.6)            \\
- Privacy \& Security                   & 80(9.8)            & 238(5.3)            & 48(7.7)            &                    & 131(14.3)          & 160(7.6)           \\
- Sensor Device                         &                    &                     & 42(6.8)            &                    & 201(22.0)          & 96(4.5)            \\
- Image Identification                  &                    & 195(4.3)            & 74(11.9)           &                    & 131(14.3)          & 147(7.0)           \\
- Hardware \& Battery                   &                    &                     &                    &                    &                    & 175(8.3)           \\
\textbf{3. Medical Issues}              & \textbf{245(30.1)} & \textbf{780(17.3)}  & \textbf{72(11.6)}  & \textbf{904(23)}   & \textbf{0(0.0)}    & \textbf{469(22.3)} \\
- Covid                                 &                    & 324(7.2)            &                    & 173(4.4)           &                    & 93(4.4)            \\
- Cancer                                & 52(6.4)            &                     & 28(4.5)            & 138(3.5)           &                    &                    \\
- Hospital (Cure)                       & 60(7.4)            &                     &                    &                    &                    & 118(5.6)           \\
- Blood                                 & 57(7.0)            &                     &                    &                    &                    & 84(4.0)            \\
- Smoking                               &                    &                     &                    & 127(3.2)           &                    &                    \\
- Clinical Consultation                 & 76(9.3)            & 271(6.0)            & 44(7.1)            & 158(4.0)           &                    &                    \\
- Detecting Model                       &                    &                     &                    & 148(3.8)           &                    & 71(3.4)            \\
- Depression \& Mental   Disorder       &                    & 185(4.1)            &                    & 160(4.1)           &                    & 103(4.9)           \\
\textbf{4. Population Demographics}     & \textbf{124(15.3)} & \textbf{669(14.8)}  & \textbf{45(7.3)}   & \textbf{459(11.7)} & \textbf{0(0.0)}    & \textbf{0(0.0)}    \\
- Family                                &                    & 232(5.1)            &                    &                    &                    &                    \\
- Child                                 &                    &                     &                    & 127(3.2)           &                    &                    \\
- Youth                                 & 60(7.4)            &                     &                    & 168(4.3)           &                    &                    \\
- Older                                 & 64(7.9)            & 158(3.5)            & 45(7.3)            &                    &                    &                    \\
- Gender                                &                    & 279(6.2)            &                    & 164(4.2)           &                    &                    \\
\textbf{5. Social Phenomenon}           & \textbf{57(7.0)}   & \textbf{1056(23.4)} & \textbf{105(16.9)} & \textbf{895(22.8)} & \textbf{150(16.4)} & \textbf{427(20.1)} \\
- Industry                              &                    & 280(6.2)            &                    &                    &                    & 87(4.1)            \\
- Area                                  & 57(7.0)            &                     &                    & 125(3.2)           &                    &                    \\
- Social Media                          &                    &                     & 39(6.3)            & 154(3.9)           &                    & 115(5.4)           \\
- Community                             &                    & 310(6.9)            &                    & 183(4.7)           & 150(16.4)          &                    \\
- Futurology                            &                    &                     & 66(10.6)           & 127(3.2)           &                    & 85(4.0)            \\
- Online Communication                  &                    & 227(5.0)            &                    &                    &                    &                    \\
- Education                             &                    & 239(5.3)            &                    & 120(3.1)           &                    & 140(6.6)           \\
- Innovation \& Ethics                  &                    &                     &                    & 186(4.7)           &                    &                    \\
\textbf{6. Healthcare}                  & \textbf{0(0.0)}    & \textbf{340(7.5)}   & \textbf{80(12.9)}  & \textbf{222(5.7)}  & \textbf{0(0.0)}    & \textbf{118(5.6)}  \\
- Physical Activity                     &                    & 196(4.3)            & 39(6.3)            & 140(3.6)           &                    & 118(5.6)           \\
- Daily Care (prevention)               &                    & 144(3.2)            & 41(6.6)            & 82(2.1)            &                    &                    \\
\textbf{Total Number of Articles}       & 813                & 4514                & 620                & 3933               & 915                & 2115               \\ \hline
\end{tabular}
\label{tab:table6}
\end{table}

Forty percent of the articles in the medicine field focused on IT in the first half. However, the number of articles dealing with medical issues soared during the second half, also addressing journal article methodology and social phenomena in the second half. Covid-19, smoking, and mental disorders have particularly become prominent medical issues. In the second half, area, community, education, innovation, and ethics were specialized topics in social phenomena. Generally, the topics in the second half were from the perspectives of cure and treatment.

In the case of EECS, articles mostly focused on IT in the first half. The only other topic not involved in IT was community, and the proportion of articles focusing on IT was more than 80 percent. However, the EECS is the field that changed the most in composition over time. The EECS started to look at medical issues like Covid-19 and mental disorders. It also diversified its topic to social phenomena, such as industry, social media, futurology, and education. Nevertheless, papers on IT have been emphasized and diversified in the second half. For instance, traditional topics, such as image or sensor analysis through computers, expanded to involve topics such as hardware, batteries, systems, and architecture at the technical level.

Generally, public health showed considerable interest in IT and medical issues in the first half, whereas it became more focused on social phenomena and healthcare in the second half. Papers in the medicine field also showed considerable interest in IT in the first half. However, they started to focus on their primary domain, which is medical issues, and showed substantial inclination toward social phenomena in the second half. Finally, the articles in the EECS showed the most significant fluctuations. They remained in the IT field in the first half but started to diversify in the second half toward social phenomena and medical issues. However, interest in IT continued to dominate.

Overall, all three fields raise interest in social phenomena and healthcare in the second half. In particular, topics such as Covid-19, depression and mental disorders, education, and physical activity appeared for the first time in the second half of all three fields. Even though the proportion of themes fluctuated in each LDA model, the number of articles in each theme increased in the second half. Digital health had established more distinguished areas in the second half.

\subsubsection{Homogeneity Test}

\begin{table}[!hbt]
\caption{Theme distribution of public health in the 1st and 2nd half}
\centering
\begin{tabular}{cccc}
\hline
Topics of public health &
  \begin{tabular}[c]{@{}c@{}}2012-2016\\ Frequency (Percent)\end{tabular} &
  \begin{tabular}[c]{@{}c@{}}2017-2022\\ Frequency (Percent)\end{tabular} &
  Total \\ \hline
Journal Article Methodology & 59(7.3)   & 850(18.8)  & 909(17.1)  \\
Information Technology      & 328(40.3) & 819(18.1)  & 1147(21.5) \\
Medical Issues              & 245(30.1) & 780(17.3)  & 1025(19.2) \\
Subject                     & 124(15.3) & 669(14.8)  & 793(14.9)  \\
Social Phenomenon           & 57(7.0)   & 1056(23.4) & 1113(20.9) \\
Healthcare                  & 0(0.0)    & 340(7.5)   & 340(6.4)   \\
Total                       & 813(100)  & 4514(100)  & 5327(100)
\\ \hline
\end{tabular}
\label{tab:table7}
\end{table}

\begin{table}[!hbt]
\caption{Table of Homogeneity Test for each comparison}
\centering
\begin{tabular}{ccc}
\hline
Comparison                                             & Bonferroni adjusted P-value             \\ \hline
\begin{tabular}[c]{@{}c@{}}Public health: 1st half vs 2nd half\end{tabular}                 & \textless{}0.05 \\  
\begin{tabular}[c]{@{}c@{}}Medicine: 1st half vs 2nd half\end{tabular}                     & \textless{}0.05 \\ 
\begin{tabular}[c]{@{}c@{}}EECS: 1st half vs 2nd half\end{tabular}                   & \textless{}0.05 \\ 
\begin{tabular}[c]{@{}c@{}}1st half: Public health vs Medicine vs EECS\end{tabular}  & \textless{}0.05   \\ 
\begin{tabular}[c]{@{}c@{}}1st half\: Public health vs Medicine\end{tabular}           & \textless{}0.05 \\ 
\begin{tabular}[c]{@{}c@{}}1st half: Public health vs EECS\end{tabular}                 & \textless{}0.05 \\  
\begin{tabular}[c]{@{}c@{}}1st half: Medicine vs EECS\end{tabular}      & \textless{}0.05 \\   
\begin{tabular}[c]{@{}c@{}}2nd half: Public health vs Medicine vs EECS\end{tabular} & \textless{}0.05 \\ 
\begin{tabular}[c]{@{}c@{}}2nd half: Public health vs Medicine\end{tabular}     & \textless{}0.05 \\ 
\begin{tabular}[c]{@{}c@{}}2nd half: Public health vs EECS\end{tabular}       & \textless{}0.05 \\ 
\begin{tabular}[c]{@{}c@{}}2nd half: Medicine vs EECS\end{tabular}           & \textless{}0.05 \\ \hline
\end{tabular}
\label{tab:table8}
\end{table}

A homogeneity test was conducted to compare whether there was a difference in topic composition between each field and period. The number of published articles was compared from the perspective of themes. In other words, the distributions of themes in the different LDA topic models were compared. The results are presented in Tables 7 and 8. Table 7 compares the first and second halves in public health, while Table 8 reveals the overall results. Fisher’s exact test was applied since there were cells containing less than 5 samples and Bonferroni correction was applied to adjust p-values for multiple comparisons. The results show that public health, medicine, and EECS changed in the composition of topics in the second half, and the three domains had different topic compositions.

Since F-test only verifies whether the entire group is equally distributed, this paper applied every T-test to look for any fluctuation between time periods and themes. According to the T-test, for both the first and second halves, there was no similarity in topic composition between any of the LDA topic models, as shown in Tables 7 and 8.

\section{Discussion}
\label{sec:discussion}
\subsection{Findings}
Among 15950 digital health-related papers published on the WoS, most belonged to 3 domains: public health, medicine, and EECS. There were apparent differences in the three academic domains in the first half that the number of published papers differed. However, the public health and medicine domains had a similar composition from a qualitative perspective since they focused on IT and medical issues in the first half. 

The specialization of each domain occurred in the second half. Specifically, in the case of EECS, technical issues were intensively addressed from a methodological point of view, while public health dealt with social phenomena and medicine focused on medical issues in the second half. In particular, the number of papers dealing with social phenomena in public health increased more than 20 times in the second half compared with the first half. Overall, the proportion of papers representing their original domains was small in the first half but increased significantly in the second half. In short, academic fields have become more distinguished and specialized over time.

As a result of applying LDA topic modeling, all six LDA topic models had different topics but showed a limited range of themes. These were journal article methods, IT, medical issues, population demographics, social phenomena, and healthcare. In the case of public health, the diversification of topics appeared in the second half in terms of social phenomena and healthcare, mainly focusing on community, online communication, education, physical activity, and daily care. The range of interest in medicine has expanded from IT to medical issues. In terms of treatment and cure, they have begun to broaden their interests beyond IT issues. In the case of the EECS, the field has recently expanded to focus on medical issues and social phenomena, while articles in the past only paid attention to methodological issues. EECS was clearly distinguished from public health and medicine in the first half since the latter two focused on IT and medical issues. As time passed, social phenomena and healthcare became popular topics, with topics such as Covid-19, depression, education, and physical activities becoming common. 

By 2021, the number of papers written in public health, medicine, and EECS was 6454, 5481, and 4015, respectively. Of these, 2155 belonged to both public health and medicine. This is a surprising number considering that the number of papers belonging to both public health and EECS was 46, and the number of papers belonging to both EECS and medicine was 254.

\subsection{Limitations}
The trends in digital health articles are examined only based on WoS, with no consideration of other databases. While WoS is undoubtedly representative of the research trends, digital health is a broad field covered by a more significant number of databases such as Pub Med, Scopus, and Embase.

In the case of LDA topic modeling, there were fewer qualitative analysis elements. Hence, topic modeling can be improved by introducing qualitative analysis theory. Since all documents cannot be qualitatively analyzed, the overall topics can be extracted using LDA topic modeling; however, specifying them requires a professional interpretation of each domain.

\subsection{Conclusions}
Amid a significant surge of interest in digital health, articles on digital health are primarily found in public health, medicine, and EECS. The topics in each domain differed over time. In the first half of the last decade, there were no significant differences between the topics of public health and medicine. 

EECS was limited to technological topics, while public health tended to be interested in social phenomena, particularly community, online communication, and education. In the case of medicine, the introduction of emotional disorders and therapeutic models has expanded the scope of medical care and increased interest in IT techniques. The EECS has consistently expanded and focused on technical issues using computers, but it has gradually come to focus on social and medical issues. 

With Covid-19 becoming a dominant issue recently, digital health has come to be strongly related to depression and mental disorders, education, and physical activity, with articles on these topics appearing in the second half in all fields. The scope of digital health research is expanding and its composition fluctuates. Digital health researchers should emphasize and understand this trend. There was no bias in topic composition among the three domains, but other domains such as kinesiology or psychology would contribute to digital health research in the future. 

In addition, it is also necessary to advance the clustering technique of topics to include qualitative research to reflect future global research trends better. In future studies, we plan to compare and analyze various topic-modeling techniques to derive better models and conduct research that can specify the scope of digital health. Finally, exploring papers on expanded topics that reflect people's needs for digital health over time will be necessary for future research.


\section{Conflicts of interest}
\label{sec:conflicts of interest}
None declared.

\section{Abbreviations}
\label{sec:abbreviations}
\begin{itemize}
\item LDA: Latent dirichlet allocation 
\item EECS: Electrical engineering and computer science 
\item Covid-19: Coronavirus disease of 2019 
\item IT: Innovation technology 
\item PC: Principal Component 
\item SCIE: The Science Citation Index Expanded
\end{itemize}




\end{document}